\documentclass[a4paper, 10pt, conference]{ettc-conf}

\IEEEoverridecommandlockouts
\usepackage[utf8]{inputenc}
\usepackage{lmodern}
\usepackage{times}
\usepackage{amsmath,amssymb,amsfonts}
\usepackage{algorithmic}
\usepackage{graphicx}
\usepackage{xcolor}
\usepackage{float}
\usepackage{schemabloc}
\usepackage{circuitikz}
\usepackage{tikz}
\usepackage{hyperref}
\usepackage{todonotes}
\usepackage[numbib]{tocbibind} 
\usepackage[labelfont=bf]{caption}
\usepackage{subcaption}
\usepackage{fancyhdr}

\makeatletter
 \let\old@ps@headings\ps@headings
 \let\old@ps@IEEEtitlepagestyle\ps@IEEEtitlepagestyle
 \def\conffooter#1{%
 \def\ps@headings{%
 \old@ps@headings%
 \def\@oddfoot{\strut\hfill#1\hfill\strut}%
 \def\@evenfoot{\strut\hfill#1\hfill\strut}%
 }%
 \def\ps@IEEEtitlepagestyle{%
 \old@ps@IEEEtitlepagestyle%
 \def\@oddfoot{\strut\hfill#1\hfill\strut}%
 \def\@evenfoot{\strut\hfill#1\hfill\strut}%
 }%
 \ps@headings%
 }
 \makeatother

\conffooter{%
 \textit{\small ETTC 2023– European Test \& Telemetry Conference} 
 }

\hypersetup{
colorlinks= true,
    citecolor = blue ,
    linkcolor= red,
}

\usetikzlibrary{positioning,
                intersections,
                backgrounds,
                shapes.geometric,
                fit,
                arrows,
                decorations.shapes,
                decorations.markings}

\tikzset{%
  block/.style      = {draw, thick, rectangle, minimum height = 3em, minimum width = 3em, inner sep=2mm},
  gain/.style       = {draw, thick, isosceles triangle, isosceles triangle apex angle=45,scale=1,sharp corners, anchor=center},
  input/.style      = {coordinate}, 
  output/.style     = {coordinate},  
  add/.style n args={4}{
    minimum width=8mm,
    path picture={
        \draw[black] 
            (path picture bounding box.south east) -- (path picture bounding box.north west)
            (path picture bounding box.south west) -- (path picture bounding box.north east);
        \node at ($(path picture bounding box.south)+(0,0.13)$)     {\scriptsize #1};
        \node at ($(path picture bounding box.west)+(0.13,0)$)      {\scriptsize #2};
        \node at ($(path picture bounding box.north)+(0,-0.13)$)    {\scriptsize #3};
        \node at ($(path picture bounding box.east)+(-0.13,0)$)     {\scriptsize #4};
        }
    },
 chn1_block/.style={draw=black, line width=0.5pt, fill=white,
            inner ysep=3mm,inner xsep=5mm, rectangle, rounded corners},
 chnN_block/.style={draw=black, line width=0.5pt, fill=gray!20,
            inner ysep=3mm,inner xsep=5mm, rectangle, rounded corners}              
}

\begin{document}


\title{Real-Time Evaluation of an Ultra-Tight GNSS/INS Integration Based on Adaptive PLL Bandwidth}
\author{\IEEEauthorblockN{G. Pages, B. Priot, G. Beaugendre}
\IEEEauthorblockA{\\ ISAE-SUPAERO, University of Toulouse, 10 avenue Edouard Belin, 31055 Toulouse, France}
}

\maketitle

\begin{abstract}
In this contribution, we propose a GNSS/INS ultra-tight coupling in which the GNSS receiver architecture is based on a vector tracking loop type architecture. In the proposed approach, the phase lock loop bandwidth is adapted according to the inertial navigation system information. The latter has the advantage to be easily implementable on a System-on-Chip component such as an FPGA (Field-Programmable Gate Arrays),  and can be implemented with minor modifications on an existing GNSS receiver platform. Moreover, compared to classical vector-based solutions, the proposed architecture decodes the navigation message in the loop, without the need to run scalar loops in parallel or having to store pre-downloaded ephemeris data. This  architecture therefore does not increase the area occupied on the FPGA and does not use additional resources for storage. The proposed GNSS receiver architecture uses GPS L1/C and Galileo E1 signals and is composed of one acquisition module and 16 tracking channels (8 GPS and 8 Galileo) which are implemented within a FPGA (Zynq-Ultrascale). 
\end{abstract}


\section{Introduction}

In a desire to make transportation increasingly autonomous, the need for a robust and precise positioning system is essential. Global Navigation Satellite Systems (GNSS) is typically the technology of choice to provide positioning, velocity and timing (PVT) information. Standard GNSS receivers are based on scalar tracking loops (STL) in order to track the signal of each satellite in view independently \cite{Kaplan2005}. Although, in harsh environments, such as urban canyons, the GNSS signals are strongly degraded by multipaths, signal attenuation and non-line-of-sight phenomena which highly impacts the stability of the tracking loops and thus affects the robustness and the precision of the PVT solution. 

Other GNSS receiver architectures are proposed within the literature which overcome some of the weakness of a STL architecture. They are based on vector tracking loops (VTL) where each tracking loop is intimately related with the others through an Extended Kalman Filter (EKF). The major difference between STL and VTL receivers is that, in the latter, the navigation filter sends back information to the tracking loops: classically, the code and carrier NCOs (Numerically Controlled Oscillators) are controlled through the state estimated by the EKF \cite{Haotian2021}, \cite{Serant2012}. 
Moreover, enhancing the VTL with inertial measurement unit (IMU) information through an inertial navigation system (INS), in an ultra-tightly coupled manner, allows to account for local receiver dynamics. Thereby it is possible to narrow down the bandwidth of phase lock loop (PLL) thus improving the resilience to noise and the stability of the tracking loops. When it comes to VTL architectures, several approaches can be considered, namely vector delay lock loops (VDLL), vector frequency lock loops (VFLL) and vector delay and frequency lock loops (VDFLL) \cite{Rongjun2021}, \cite{Dietmayer2020-PLANS}, \cite{Elghamrawy2020}. 

In this paper we consider VDFLL architectures. They are mainly based on either centralized or decentralized architectures. In a centralized architecture, the loop filter has been withdrawn from the tracking loops and the outputs of either the correlators or the discriminators are directly used as inputs of the navigation filter. Although centralized architectures are computationally expensive and present nonlinearities which need to be addressed \cite{Babu2009}. As for the decentralized architecture, each tracking channel has a local filter, mostly based on a Kalman Filter (KF) type estimator. The state estimated by the local filter of each channel is viewed as a measurement by the navigation filter. Works on VDFLL architectures, such as in \cite{Rongjun2021}, \cite{Elghamrawy2020}, \cite{Sousa2014}, show good results in specific environments. Unfortunately the real-time hardware implementation, such as on FPGAs (Field Programmable Gate Array), remains complex and data synchronization and calculation time management between the EKF and the tracking loops is not trivial to handle \cite{Dietmayer2020-ION}. In addition, most of the VDFLL architectures in literature need an STL architecture to run in parallel in order to initialise the VTL architecture which, increasing thus the area occupied on an FPGA as well as the complexity of the full system.

In this paper  we propose an alternative to VDFLL architectures, seeking the trade-off between performance and ease of implementation on FPGA components. Our architecture is based on an existing GNSS STL receiver which requires few modifications, and, nevertheless, seems to be just as efficient as a classical VDFLL architecture. 

The rest of the paper is organized as follows: Section 2 presents the ultra-tightly coupled decentralized VDFLL. In Section 3 the proposed ultra-tight architecture (UT-ALFA) is described. Section 4 details the design and the implementation of the UT-ALFA on FPGA. Experimental results are then presented in Section 5 followed by a conclusion in Section 6.

\section{Ultra-tightly coupled decentralized VDFLL}
\begin{figure}[t]
    \centering
\begin{tikzpicture}[scale=0.65, transform shape, auto, thick, node distance=2cm, >=triangle 45]
    \node at (0,0) [input] (fe) {};
    \node at (9,-5.2) [output] (cpgfeedback) {};
    \node at (7.75,-4) [input] (nco-og-1) {};
    \node at (7.75,-2) [input] (nco-og-2) {};

    \node at (2,0) [block, minimum height=1cm,text width=2cm, align = center, font=\normalsize] (corr) {Correlators};
    \node at (6,0) [block, minimum height=1cm,text width=2cm, align = center, font=\normalsize] (discri) {Discriminators};
    \node at (6,-2) [block, minimum height=1cm,text width=2.15cm, align = center, font=\normalsize] (loop_filter) {Local estimator};
    
    \node at (9,-2) [block, minimum height=1cm,text width=1cm, align = center, font=\normalsize] (obs_gen) {OG};
    
    \node at (6,-4) [block, minimum height=1cm, text width=2.15cm, align = center, font=\normalsize] (nco) {Code \& Carrier NCOs};
    \node at (2,-4) [block, minimum height=1cm,text width=1.5cm, align = center, font=\normalsize] (replica) {Signal generator};
    \node at (12,-3) [block, minimum height=3cm, text width=1.6cm, align = center, font=\normalsize] (nav_filter) {Navigation \\ filter};
    \node at (12,0) [block, minimum height=1.2cm, text width=1.6cm, align = center, font=\normalsize] (imu) {IMU};

    \node at (9,-4) [block, minimum height=1cm,text width=1cm, align = center, font=\normalsize] (cpg) {CPG};

    \draw[-stealth](fe) to (corr);

    \draw[-stealth](corr) -- node[above=0.1cm,midway, font=\large] {$E$, $P$, $L$} (discri);
    \draw[-stealth]([xshift=-0.8cm]discri.south) to ([xshift=-0.8cm]loop_filter.north);

    \draw[-stealth](discri) to (loop_filter);
    \draw[-stealth]([xshift=0.8cm]discri.south) to ([xshift=0.8cm]loop_filter.north);
    \draw (5.2,-0.65) node [anchor=north west, align=right, font=\large] {$\delta\Tilde{\tau}$};
    \draw (6.0,-0.65) node [anchor=north west, align=right, font=\large] {$\delta\Tilde{\varphi}$};
    \draw (6.8,-0.6) node [anchor=north west, align=right, font=\large] {$\delta \Tilde{f}_d$};

    \draw[-stealth](loop_filter.south) -- node[midway, align=right, font=\large] {$\boldsymbol \theta^\text{STL}$} (nco.north);

    \draw[-stealth]([yshift=0.3cm]loop_filter.east) -- node[above=0.5mm, midway, font=\large] {$\hat{\mathbf{x}}$}  ([yshift=0.3cm]obs_gen.west);

    \draw[-stealth]([yshift=0.3cm]nco.west) -- node[above=0.5mm, midway, font=\large] {$\tau^{\text{NCO}}$} ([yshift=0.3cm]replica.east);
    \draw[-stealth]([yshift=-0.3cm]nco.west) -- node[below=0.5mm, midway, font=\large] {$\varphi^{\text{NCO}}$} ([yshift=-0.3cm]replica.east);

    \draw[](nco.south) |- (cpgfeedback);
    \draw[-stealth](cpgfeedback) -- (cpg.south);

    \draw[]([yshift=0.3cm]nco.east) -- ([yshift=0.3cm]nco-og-1.west);
    \draw[]([yshift=0.3cm]nco-og-1.north) -- ([yshift=-0.3cm]nco-og-2);
    \draw[-stealth]([yshift=-0.3cm]nco-og-2.east) -- ([yshift=-0.3cm]obs_gen.west);

    \draw (7.75,-2.8) node[anchor=north west, align=right, font=\large] {$\tau^\text{NCO}$};
    \draw (6.8,-5.2) node[anchor=north west, align=right, font=\large] {$\tau^\text{NCO}$, $T_\text{RX}$};
    

    \draw[-stealth]([xshift=0.55cm]replica.north) to ([xshift=0.55cm]corr.south);
    \draw[-stealth]([xshift=-0.55cm]replica.north) to ([xshift=-0.55cm]corr.south);

    \draw[-stealth](obs_gen.east) -- node[above=0.5mm, midway, font=\large] {$\boldsymbol \eta$}  ([yshift=1cm]nav_filter.west);

    \draw[-stealth](nav_filter.west |- nco.east) -- node[above=0.5mm, midway, font=\large] {$\boldsymbol \zeta$} (cpg.east);

    \draw[-stealth]([yshift=-0.3cm]cpg.west) -- node[below=0.5mm, midway, font=\large] {$\boldsymbol \theta^\text{VTL}$}  ([yshift=-0.3cm]nco.east);

    \draw[-stealth](imu.south) -- node[midway,font=\large] {$\tilde{\mathbf{a}}$, $\tilde{\boldsymbol{\omega}}$} (nav_filter.north);

\end{tikzpicture}
    \caption{\textbf{Ultra-tightly coupled vector delay and frequency lock loop architecture for one tracking channel.}}
    \label{fig:vdfll-architecture}
\end{figure}
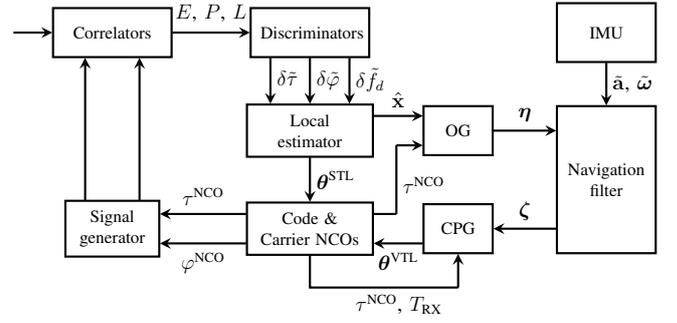

In this paper, a decentralized VDFLL architecture ultra-tightly coupled with an INS is adopted, as depicted in Fig. \ref{fig:vdfll-architecture}. Note that the acquisition phase of the receiver is not represented and will not be considered in this study. This architecture will serve as the reference vector-based receiver to which we will compare our proposed architecture and is detailed within this section. 

Here, ${\boldsymbol \theta}^{\text{STL}}$ and ${\boldsymbol \theta}^{\text{VTL}}$ are the estimated control parameters vectors. They are both two-dimensional vectors including $f_{\text{DLL}}$ and $f_{\text{PLL}}$. ${\boldsymbol \theta}^{\text{STL}}$ is only used to initialize the VDFLL architecture. Once the navigation filter is able to operate, that is the ephemeris data is decoded and that there are sufficient observations, the STL control parameter is deactivated and ${\boldsymbol \theta}^{\text{VTL}}$ is used instead. We define $\boldsymbol \eta = \{\left[ \begin{array}{cc}\tilde{\rho} & \tilde{f}_{d} \end{array} \right]_{i}\}_{i=1}^{N}$ the set of vector observations composed of the pseudorange and the Doppler frequency shift provided by each channel $i$. We also define ${\boldsymbol \zeta} = \{\left[ \begin{array}{cc}{\rho} & {\dot\rho} \end{array} \right]_{i}\}_{i=1}^{N}$ the set of updated pseudorange and pseudorange rate respectively, provided by the navigation filter to each channel in order to calculate ${\boldsymbol \theta}^{\text{VTL}}$. 

In the rest of the paper, and for the sake of notation simplicity, we are only considering one tracking channel, thus allowing to omit the subscript $i$.

\subsection{General description}
The incoming signal is processed through the tracking loops, undergoing correlation with local carrier and code replicas for a given integration period $T_{I}$. This leads to the inphase and quadrature Early, Prompt and Late correlator outputs (respectively $E$, $P$, $L$). The latter are then provided to the discriminators which determine the errors between the parameters of the replicated signal, $\{\tau^{\text{NCO}}, \varphi^{\text{NCO}}, f_d^{\text{NCO}}\}$, and the parameters of the incoming signal, $\{\tau, \varphi, f_{d}\}$. These errors can be modeled as
\begin{eqnarray}
    \delta \tau & = & \tau - \tau^{\text{NCO}}, \label{eq:delta-tau-1}\\
    \delta \varphi & = & \varphi - \varphi^{\text{NCO}}, \label{eq:delta-phi-1}\\
    \delta f_{d} & = & f_{d} - f_{d}^{\text{NCO}}. \label{eq:delta-doppler-1}
\end{eqnarray}

The discriminators outputs are used as inputs of the local estimator, modeled as a Kalman Filter (KF), which provides the control parameter vector $\boldsymbol{\theta}^{\text{STL}}$ to the NCO and the estimated state, $\hat{\mathbf{x}}= \left[ \begin{array}{ccc}\delta\hat{\tau} & \delta\hat{\varphi}  &  \delta \hat{f}_{d} \end{array} \right] ^{\top}$. The evolution model providing the state vector at an instant $k$ is based on equations \eqref{eq:delta-tau-1}-\eqref{eq:delta-doppler-1} and is  formulated as
\begin{equation}
    \mathbf{x}_k = \mathbf{F}_{k-1}\mathbf{x}_{k-1} + \mathbf{G}_{k-1} \mathbf{u}_{k-1} + \mathbf{w}_{k-1} \label{eq:ssm-f}\\
\end{equation}
where
\begin{equation*}
   \begin{array}{cc}
     \mathbf{F}_{k-1} = \begin{bmatrix} 1 & 0 & \alpha T_I\\ 0 & 1 & T_I \\ 0 & 0 & 1 \end{bmatrix} &  \mathbf{G}_{k-1} = \begin{bmatrix} -T_I & \alpha T_I \\ 0 & 0  \\ 0 & 0 \end{bmatrix},
\end{array} 
\end{equation*}
$\alpha = -{f_{\text{chip}}}/{f_c}$,  $f_{\text{chip}} = 1.023$MHz is the code frequency, $f_c$ is the carrier frequency and $\mathbf{u}_{k-1} = \begin{bmatrix} f_{\text{DLL}} & f_{\text{PLL}} \end{bmatrix}_{k-1}^{\top}$. The term $\mathbf{w}_{k-1}$ is considered to be a zero mean uncorrelated Gaussian process noise such that $\mathbf{w}_{k-1} \sim \mathcal{N}(\mathbf{0},\mathbf{Q}_{k-1})$, where $\mathbf{Q}_{k-1}$ defines the process covariance matrix.  
The observation model is expressed as
\begin{equation}
    \mathbf{z}_k  =  \mathbf{H}_k\mathbf{x}_{k} + \mathbf{n}_k, \label{eq:ssm-h}
\end{equation}
where the measurement vector is $\mathbf{z}_k = \begin{bmatrix} \delta\tilde{\tau}_k & \delta\tilde{\varphi}_k  &  \delta\tilde{f}_{d,k} \end{bmatrix}^{\top}$ and $\mathbf{n}_{k}\sim \mathcal{N}(\mathbf{0},\mathbf{R}_{k})$ is the zero mean uncorrelated Gaussian noise affecting the observations, represented by the observation covariance matrix $\mathbf{R}_{k}$. $\mathbf{H}_k = \mathbf{I}_{3\times 3}$ is the observation matrix that maps the states into the observations. The OG (Observation Generator) determines the observation vector $\boldsymbol{\eta}$ from the estimated state, $\tau^{\text{NCO}}$ and $f_{\text{PLL}}$. The navigation filter is modeled as an Extended Kalman Filter (EKF) which tightly couples the inertial measurements with the provided GNSS observations. The state vector which is estimated by the EKF compromises the 3-dimensional position and velocity, the attitude in quaternion, the 3-dimensional acceleration and rotation rates bias as well as the GNSS receiver's clock bias and clock drift. More details on the tightly coupled IMU/GNSS integration can be found in \cite{Gro2013}. From the estimated states, the EKF provides updates of the pseudorange $\rho$ and the pseudorange rate $\dot{\rho}$ of each tracked satellite. The latter are sent back to each tracking channel. Finally, the NCO control parameter vector $\boldsymbol \theta^{\text{VTL}}$ is obtained within the CPG (Control Parameter Generator) using $\boldsymbol \zeta$ and is provided to the code and carrier NCOs which supplies the signal parameters in order to generate the local replica of the signal.

\subsection{Discussion}

Although providing good tracking performances in weak signal environments compared to the STL architecture, implementing the ultra-tightly VDFLL, on a FPGA component is still a tedious task when it comes to data synchronisation and timing issues. Indeed, when it comes to implementing this architecture from an existing STL architecture on an FPGA component, it requires major modifications. Firstly, there is a synchronization operation to be carried out during the generation of the observations (OG) requiring the use of the information from the NCO and that provided by the filter which are not produced at the same time. Secondly, and most importantly, in order to control the NCOs, an additional operation is necessary to process the data coming from the navigation filter, $\boldsymbol{\zeta}$, in order to obtain the control parameters $\boldsymbol{\theta}^{\text{VTL}}$. This is realized by the CPG. Since the NCOs run at high speed compared to the navigation filter, the CPG has the delicate task of realizing the synchronization between the data provided by the navigation filter and the code NCO to produce $\boldsymbol{\theta}^{\text{VTL}}$ in real time. This process has strong timing requirements and must be implemented in hardware. Thus to carry out these modifications it requires a deep knowledge of the receiver, which can also involve a significant development cost.

%




\section{Proposed architecture}
Figures \ref{fig:alpha-architecture} and \ref{fig:alpha-lf} illustrates the proposed architecture. The difference with the previous VDFLL architecture is that we have kept the traditional second order FLL assisted PLL of the STL architecture \cite{Kaplan2005} and instead of sending information back to the NCOs, we send it directly back to the loop filter, in order to control the bandwidth of the PLL. The latter is thus adapted accordingly to the dynamic of the vehicle and, moreover, we resolve the problem related to data synchronization as it was stated in the VDFLL. We define ${\boldsymbol \xi} = \{\dot{f}_{d,i}\}_{i=1}^{N}$ as the set of Doppler frequency rates which are fed-back respectively to each channel $i$. Within the navigation filter, the Doppler frequency for a given satellite can be expressed as

\begin{equation}
\label{eq:doppler}
    f_{d} = -\frac{f_c}{c}\left( \mathbf{e}_{\rho}^\top \left(\mathbf{v}_u - \mathbf{v}_s\right) + c\delta \dot{t}_{u} - c\delta \dot{t}_{s} \right),
\end{equation}

where $f_c$ is the carrier frequency (Hz), $c$ is the speed of light (m/s), $\mathbf{v}_u$ and $\mathbf{v}_s$ are the respectively the receiver's and the satellite's velocity vectors (m/s), $c\delta \dot{t}_{u}$ is the receiver's clock drift and $c\delta \dot{t}_{s}$ is the satellite's clock drift (m/s). $\mathbf{e}_{\rho} = \left[e_x ~ e_y ~ e_z \right]^T$ is the line-of-sight unit vector. By deriving equation \eqref{eq:doppler} in function of time, the Doppler frequency rate is then

\begin{equation}
\label{eq:doppler_dot}
\dot{f}_{d} = -\frac{f_c}{c}\left( \dot{\mathbf{e}}_{\rho}^\top \left(\mathbf{v}_u - \mathbf{v}_s\right) + \mathbf{e}_{\rho}^\top \left(\mathbf{a}_u - \mathbf{a}_s\right) + c\delta \ddot{t}_{u} - c\delta \ddot{t}_{s} \right)
\end{equation}

where $\mathbf{a}_u$ and $\mathbf{a}_s$ are the respectively the receiver's and the satellite's acceleration vectors (m/s$^2$), $\dot{\mathbf{e}}_{\rho}$ is the derivative of the line-of-sight vector and $c\delta \ddot{t}_{u}$ and $c\delta \ddot{t}_{s}$ are the respectively the receiver's and the satellite's clock jerk (m/s$^2$). As the EKF has to be initialized  with an initial PVT information, the UT-ALFA receiver is designed to first use the STL. If at least four satellites are tracked with valid decoded ephemeris, the receiver switches to VTL mode by using the Doppler frequency rate provided by the EKF within the loop filter.




\begin{figure}[t]
    \centering
\begin{tikzpicture}[scale=0.65, transform shape, auto, thick, node distance=2cm, >=triangle 45]

    \node at (0,0) [input] (fe) {};
    \node at (7.75,-4) [input] (nco-og-1) {};
    \node at (7.75,-3) [input] (nco-og-2) {};

    \node at (2,0) [block, minimum height=1cm,text width=2cm, align = center, font=\normalsize] (corr) {Correlators};
    \node at (6,0) [block, minimum height=1cm,text width=2cm, align = center, font=\normalsize] (discri) {Discriminators};
    \node at (6,-2) [block, minimum height=1cm,text width=2.15cm, align = center, font=\normalsize] (loop_filter) {Loop filter};
    
    \node at (9,-3) [block, minimum height=1cm,text width=1cm, align = center, font=\normalsize] (obs_gen) {OG};

    \node at (6,-4) [block, minimum height=1cm, text width=2.15cm, align = center, font=\normalsize] (nco) {Code \& Carrier NCOs};
    \node at (2,-4) [block, minimum height=1cm,text width=1.5cm, align = center, font=\normalsize] (replica) {Signal generator};
    \node at (12,-3) [block, minimum height=3cm, text width=1.6cm, align = center, font=\normalsize] (nav_filter) {Navigation \\ filter};
    \node at (12,0) [block, minimum height=1.2cm, text width=1.6cm, align = center, font=\normalsize] (imu) {IMU};


    \draw[-stealth](fe) to (corr);

    \draw[-stealth](corr) -- node[above=0.1cm,midway,font=\large] {$E$, $P$, $L$} (discri);
    \draw[-stealth]([xshift=-0.8cm]discri.south) to ([xshift=-0.8cm]loop_filter.north);

    \draw[-stealth](discri) to (loop_filter);
    \draw[-stealth]([xshift=0.8cm]discri.south) to ([xshift=0.8cm]loop_filter.north);
    \draw (5.2,-0.65) node [anchor=north west, align=right, font=\large] {$\delta\Tilde{\tau}$};
    \draw (6.0,-0.65) node [anchor=north west, align=right, font=\large] {$\delta\Tilde{\varphi}$};
    \draw (6.8,-0.6) node [anchor=north west, align=right, font=\large] {$\delta \Tilde{f}_d$};

    \draw[-stealth](loop_filter.south) -- node[midway, anchor=east, align=right, font=\large] {${\boldsymbol \theta}_\text{STL}$} (nco.north);

    \draw[-stealth]([yshift=0.3cm]nco.west) -- node[above=0.5mm, midway, font=\large] {$\tau^{\text{NCO}}$} ([yshift=0.3cm]replica.east);
    \draw[-stealth]([yshift=-0.3cm]nco.west) -- node[below=0.5mm, midway, font=\large] {$\varphi^{\text{NCO}}$} ([yshift=-0.3cm]replica.east);

    \draw[](nco.east) -- node[below=0.1cm, anchor=north west, align=right, font=\large]{$\tau^\text{NCO}$}(nco-og-1.west);
    \draw[](nco-og-1.north) -- ([yshift=-0.3cm]nco-og-2);
    \draw[-stealth]([yshift=-0.3cm]nco-og-2.east) -- ([yshift=-0.3cm]obs_gen.west);


    \draw[-stealth]([xshift=0.55cm]replica.north) to ([xshift=0.55cm]corr.south);
    \draw[-stealth]([xshift=-0.55cm]replica.north) to ([xshift=-0.55cm]corr.south);

    \draw[-stealth]([xshift=0.4cm]loop_filter.south) |- node[anchor=south west, near end, font=\large] {$f_{d}$}  (obs_gen.west);

    \draw[-stealth](obs_gen.east) -- node[above=0.5mm, midway, font=\large] {$\boldsymbol \eta$}  (nav_filter.west);

    \draw[-stealth]([yshift=1cm]nav_filter.west) -- node[above=0.5mm, near start, font=\large] {${\boldsymbol \xi}$} (loop_filter.east);

    \draw[-stealth](imu.south) -- node[midway,font=\large] {$\tilde{\mathbf{a}}$, $\tilde{\boldsymbol{\omega}}$} (nav_filter.north);

\end{tikzpicture}
    \caption{\textbf{Ultra-tightly coupled adaptive loop filter architecture (UT-ALFA) for one tracking channel.}}
    \label{fig:alpha-architecture}
\end{figure}
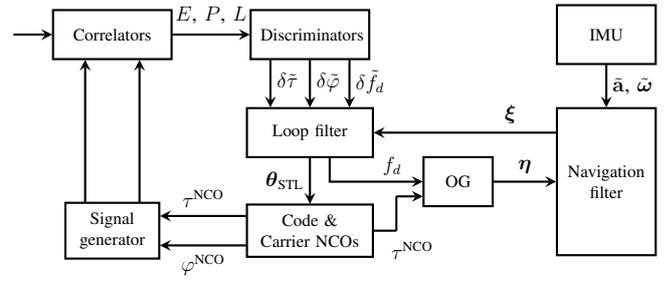
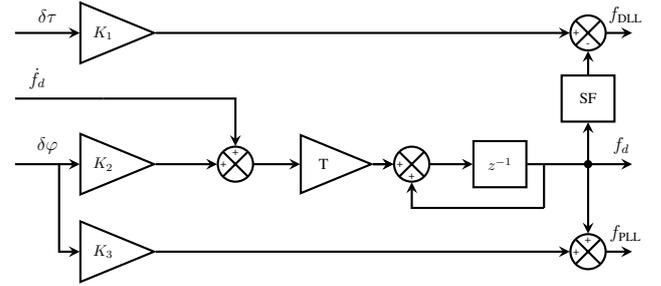
\begin{figure}[t]
    \centering
\begin{tikzpicture}[scale=0.58, transform shape, auto, thick, node distance=2cm, >=triangle 45]

\node at (0,0) [gain, text width=0.8cm, align = center, font=\normalsize] (dllgain) {$K_1$};
\node at (0,-3) [gain, text width=0.8cm, align = center, font=\normalsize] (pllgain1) {$K_2$};
\node at (0,-5) [gain, text width=0.8cm, align = center, font=\normalsize] (pllgain2) {$K_3$};
\node at (5,-3) [gain, text width=0.8cm, align = center, font=\normalsize] (int-time) {T};
\node at (11,-1.5) [block, text width=0.8cm, align = center, font=\normalsize] (scalefactor) {SF};

\node at (9,-3) [block, text width=0.8cm, align = center, font=\normalsize] (z) {$z^{-1}$};

\filldraw[black] (11,-3) circle (2pt) node[output](output-z){};

\node at (11,0) [draw,circle,add={-}{+}{}{}] (adder-dll) {};
\node at (3,-3) [draw,circle,add={}{+}{+}{}] (adder-pll1) {};
\node at (11,-5) [draw,circle,add={}{+}{+}{}] (adder-pll2) {};
\node at (7,-3) [draw,circle,add={+}{+}{}{}] (adder-int) {};

\node at (-2,0) [input] (dtau){};

\node at (-2,-1.5) [input] (fdot1){};
\node at (0,-1.5) [input] (fdot2){};

\node at (-2,-3) [input] (dphi1){};
\node at (-1,-3) [input] (dphi2){};

\node at (12,0) [output] (fdll){};
\node at (12,-3) [output] (fd){};
\node at (12,-5) [output] (fpll){};

\node at (10,-3) [output] (fbZ1){};
\node at (10,-4) [output] (fbZ2){};

\draw[-stealth](dtau) -- node[above=0.1cm,midway,font=\large] {$\delta\tau$} (dllgain);
\draw[](fdot1) -- node[above=0.1cm, near start,font=\large] {$\dot{f}_d$} (fdot2);
\draw[-stealth](fdot2) -| (adder-pll1);
\draw[-stealth](dphi1) -- node[above=0.1cm,midway,font=\large] {$\delta\varphi$} (pllgain1);
\draw[-stealth](dphi2) |- node[above=0.1cm,midway,font=\large] {} (pllgain2);

\draw[-stealth](dllgain) -- (adder-dll);
\draw[-stealth](pllgain1) -- (adder-pll1);
\draw[-stealth](pllgain2) -- (adder-pll2);

\draw[-stealth](adder-pll1) -- (int-time);
\draw[-stealth](int-time) -- (adder-int);
\draw[-stealth](int-time) -- (adder-int);
\draw[-stealth](adder-int) -- (z);

\draw[](fbZ1) -- (fbZ2);
\draw[-stealth](fbZ2) -| (adder-int);

\draw[](z) -- (output-z);
\draw[-stealth](output-z) -- (adder-pll2);
\draw[-stealth](output-z) -- (scalefactor);
\draw[-stealth](scalefactor) -- (adder-dll);

\draw[-stealth](adder-dll) -- node[above=0.1cm, near end,font=\large] {${f}_\text{DLL}$}(fdll);
\draw[-stealth](output-z) -- node[above=0.1cm, near end,font=\large] {${f}_d$}(fd);
\draw[-stealth](adder-pll2) -- node[above=0.1cm, near end,font=\large] {${f}_\text{PLL}$}(fpll);

\end{tikzpicture}
    \caption{\textbf{UT-ALFA loop filter for one channel: the FLL-assist from the classical loop filter has been replaced by the Doppler frequency rate.}}
    \label{fig:alpha-lf}
\end{figure}


Compared to the previous VDFLL architecture, the proposed architecture is simpler to implement since the synchronization between the outputs of the navigation filter and the NCO of each tracking channel is no longer required.

\section{System Design and Implementation}
\label{sec:gnss-fpga}

\begin{figure}[t]
    \centering
    \includegraphics[width=0.45\textwidth]{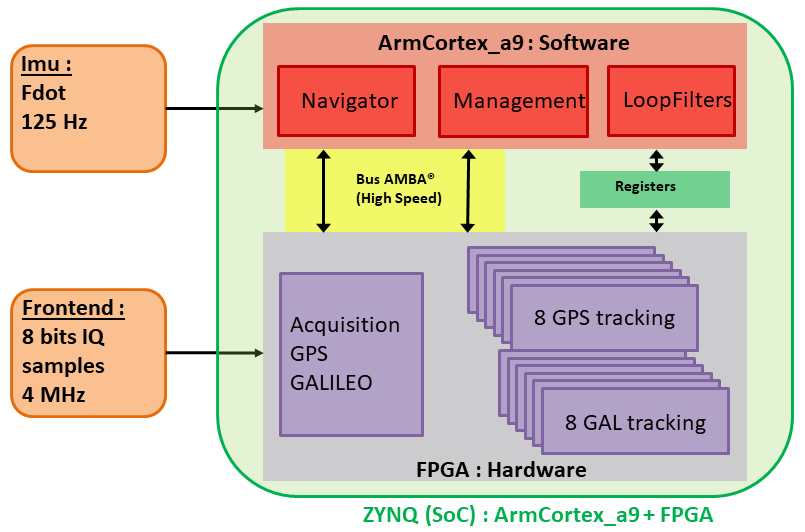}
    \caption{\textbf{ISAE-SUPAERO UT-ALFA architecture on Zynq.}}
    \label{fig:UT-APLHA architecture}
\end{figure}

The UT-ALFA was implemented on the ZCU102 (Xilinx UltraScale), as shown in Fig. \ref{fig:UT-APLHA architecture}. It consists of a fully customizable bi-constellation GNSS receiver (GPS and Galileo) from which we are able, from baseband signal, to acquire and track satellites in view, decode the navigation data and estimate the PVT (Position Velocity Time) solution.
The FPGA receives the signal samples, provided by a radio frequency front-end (FE). A data handler is responsible for dispatching the incoming signal to the appropriates processing modules, i.e. acquisition and tracking. The acquisition module is fully implemented in hardware, while the tracking channels have been partitioned so that the correlators, discriminators, the NCOs and the navigation data decoders have been integrated in hardware and the loop filter on the ARM processor. As for the navigation filter, it was implemented in Matlab in order to provide the pre-processed values of ${\boldsymbol \xi} = \{\dot{f}_{d,i}\}_{i=1}^{N}$ (Fdot in Fig. \ref{fig:alpha-architecture}) to the loop filters.


The implementation was realized through a co-design and high level synthesis tool, SpaceStudio developped by SpaceCodesign. The latter is a framework allowing to develop applications to speed-up performance using CPU and FPGA without knowing the underlying hardware infrastructure of these technologies. From a methodology point of view, using this co-design software, we performed architecture exploration in simulation to do hardware/software partitioning, events synchronization and evaluate timing constraints. With the same source code we were able to generate the software part running on linux OS, the application binaries and the drivers. We also generated HDL (Hardware Description Language) IP functions (namely acquisition and tracking channels) from C++ code using High Level Synthesis tool (Vivado HLS).

\section{Experiments}
The implementation of a standard GNSS receiver (STL) and the ultra-tightly coupled VDFLL was done in Matlab. The proposed architecture, UT-ALFA, was implemented in Matab as well as on the FPGA as detailed in section \ref{sec:gnss-fpga}. We have conducted experiments in order to evaluate our approach with both simulated and real GNSS and IMU data. The experimental set-ups and the results are presented hereafter, comparing the performance of the STL, the VDFLL and the UT-ALFA architectures in a dynamic scenario and degraded carrier-to-noise ratio (C/N0) environment. Two VDFLL were evaluated (VDFLL-1 and VDFLL-2). They differ by the values of the process noise matrix, $\mathbf{Q}$, of their local estimator: $\sigma^2_{\delta f_d} = 6.4\times 10^{-3}$ Hz$^2$ for VDFLL-1 and $\sigma^2_{\delta f_d} = 6.4\times 10^{-5}$ Hz$^2$ for VDFLL-2. The variances for the measurements were defined using the following relationship
\begin{equation}
    \sigma^{2}_{\delta x} =  \sigma^{2}_{\delta x_0}10^{-(\text{C/N0})/10},
\end{equation}

where $\delta x$ represents $\delta \tau$, $\delta \varphi$ and $\delta f_d$. We have set $\sigma^2_{\delta \tau_0} = 62.5$ chips$^2$, $\sigma^2_{\delta \varphi_0} = 7.124$ cycles$^2$ and $\sigma^2_{\delta f_{d_0}} = 4.45 \times 10^5$ Hz$^2$. In addition, the PLL bandwidth ($K_2$ and $K_3$ from Fig. \ref{fig:alpha-lf}) of the STL architecture was set to 10 Hz and that of the UT-ALFA to 3 Hz.


\subsection{Simulated data scenario}
\label{ssection:exp-sim}
\begin{figure}[t]
    \centering
    \includegraphics[width = 0.38\textwidth]{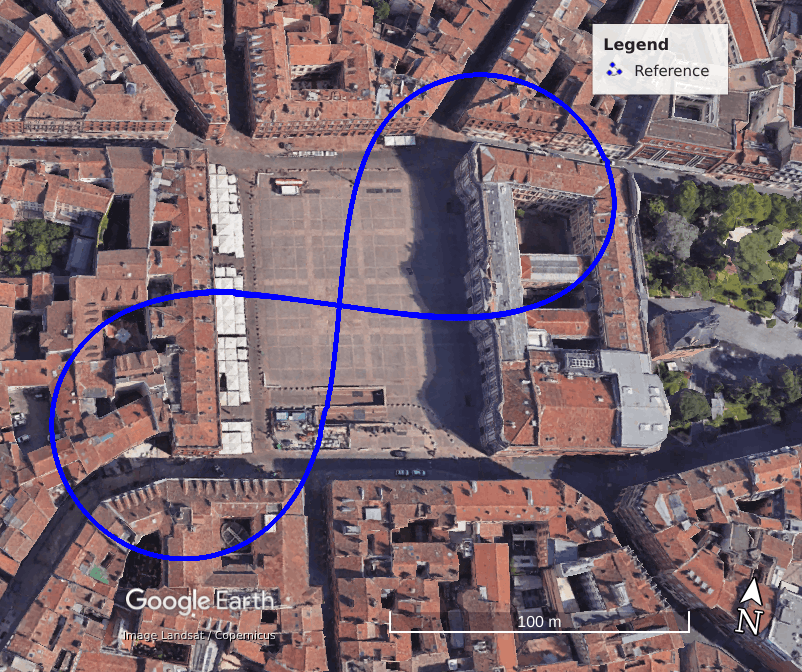}
    \caption{\textbf{Simulated trajectory visualized with Google Earth.}}
    \label{fig:sim-traj}
\end{figure}

\begin{figure}[t]
    \centering
    \includegraphics[width = 0.45\textwidth]{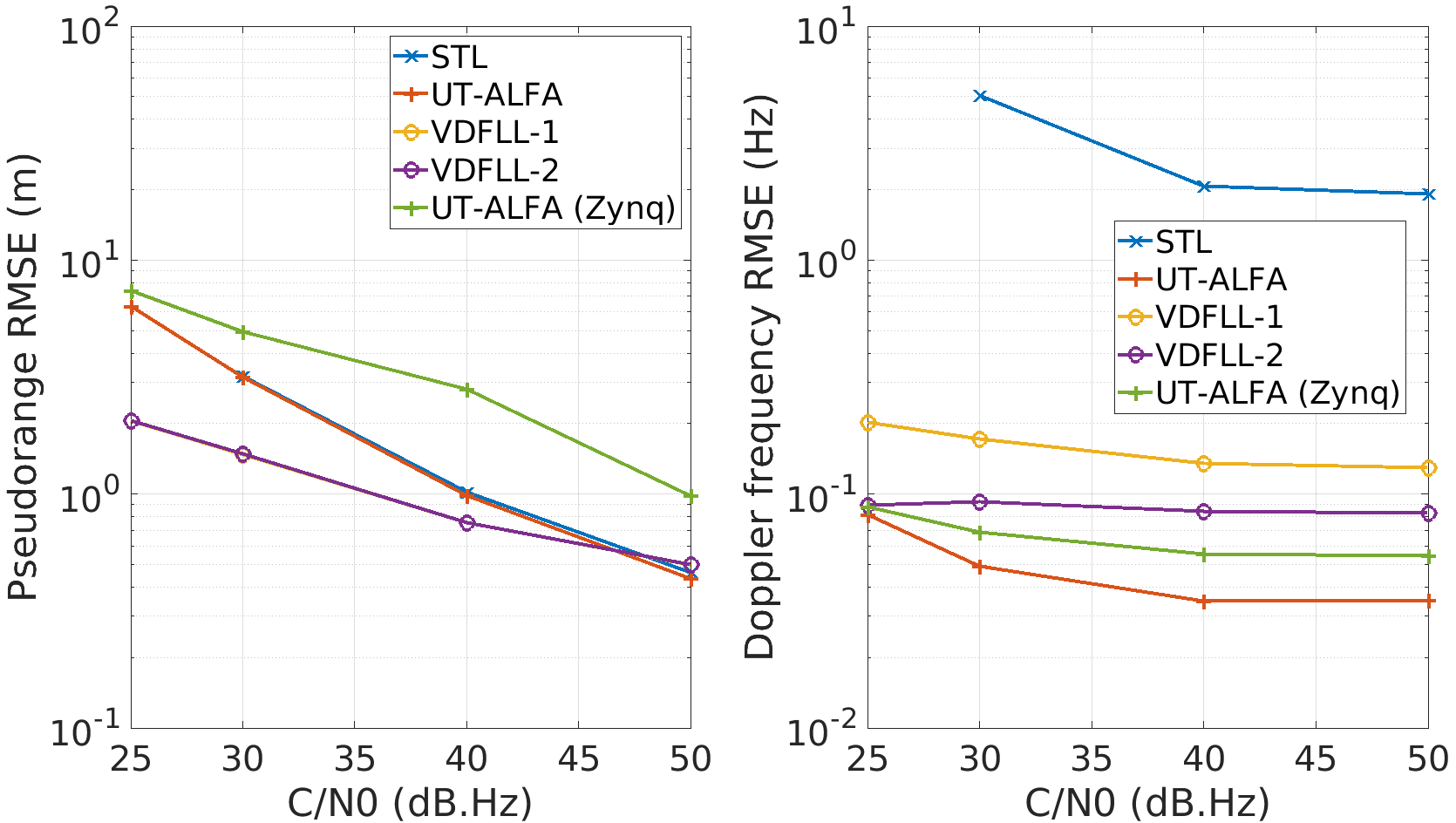}
    \caption{\textbf{Satellite 19 (elevation 28.5$^{\circ}$): Pseudorange (left) and Doppler frequency (right) RMSE relative to C/N0.}}
    \label{fig:rmse-sat19}
\end{figure}
\begin{figure}[h!]
    \centering
    \includegraphics[width = 0.45\textwidth]{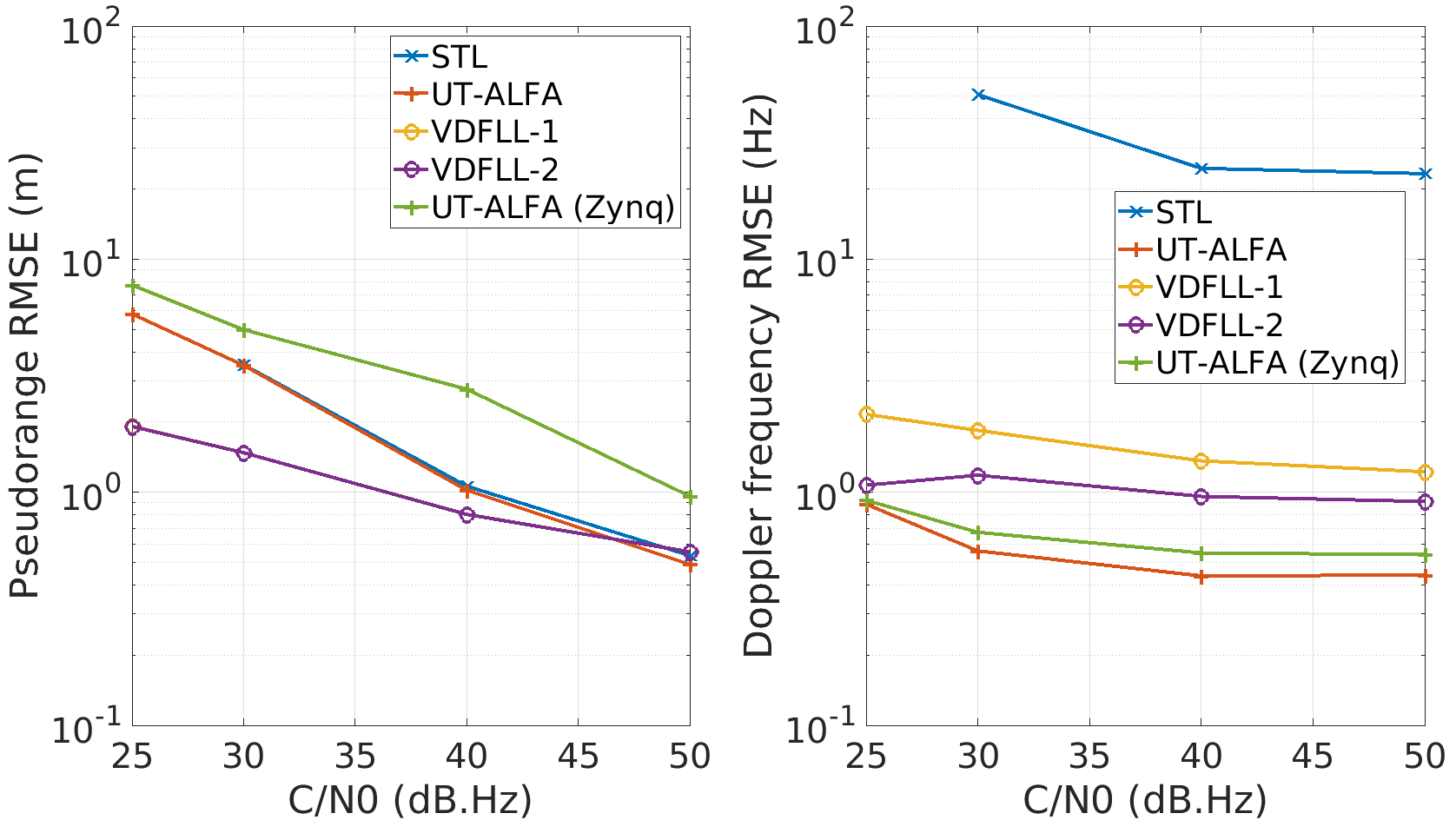}
    \caption{\textbf{Satellite 29 (elevation 11$^{\circ}$): Pseudorange (left) and Doppler frequency (right) RMSE relative to C/N0.}}
    \label{fig:rmse-sat29}
\end{figure}

The simulated scenario is illustrated in Fig \ref{fig:sim-traj} and represents the flight of a drone above the city center of Toulouse, France. We considered an open-sky view without any variation of the number of visible satellites. This simulation was done using our home-made trajectory simulator, which generates IMU data (acceleration and rotation rates) at 100 Hz. The GNSS combined 8-bit in-phase and quadrature samples were generated at a sampling frequency of 4 MHz. The GNSS signal included only the GPS constellation, compromising 8 satellites (SV\# 2, 6, 12, 17, 19, 24, 25, 29). The observables (pseudorange, Doppler frequency and C/N0) were provided at a rate of 10 Hz. 
The trajectory started with a stationary phase of 10 s. Then it accelerated in order to realize six loops of an eight-shaped figure, with several accelerations, decelerations and sharp turns, in which the maximum acceleration and velocity in the turns reached 10 m/s$^2$ and 30 m/s respectively. Once the six loops were achieved, the drone ended with another stationary phase of 10 s. During the test run, we evaluated the performances of the receivers with four different C/N0 values: 50 dB.Hz, 40 dB.Hz, 30 dB.Hz and 25 dB.Hz. During each test run, at 40 s of the simulation, the C/N0 of all the satellites was gradually degraded from 50 dB.Hz to one of the targeted C/N0 value mentioned before.

The results are given in terms of pseudorange and Doppler frequency RMSE (Root Mean Square Error) versus the C/N0 and are depicted in Fig. \ref{fig:rmse-sat19}-\ref{fig:rmse-sat29}  for satellites \# 19 and 29 (each satellite having a different elevation). The RMSE values were obtained by comparing the parameters of interests of the receivers to a reference solution for the different C/N0 values. The RMSE is represented in a logarithmic scale. We chose to show only 2 out of the 8 satellites, but the analysis of the results which  follows is the same for all the satellites. Concerning the RMSE on the pseudorange, the results show that globally the two VDFLL have the best performances and that the UT-ALFA, when compared to the STL architecture, does not lose track at low C/N0 values ($<$ 30 dB.Hz), thus still allowing to generate pseudorange measurements. Moreover, these results show that taking into account the acceleration within the receiver has no direct impact on the propagation delay estimation whatever the C/N0. On the other hand, this is different for the Doppler frequency. As we can see, the UT-ALFA fairly outperforms the VDFLLs and the STL architecture by far, with a Doppler frequency error less than 0.1 Hz  whatever the C/N0, except for satellite 29 which has a low elevation. Finally, the UT-ALFA on Zynq still shows lower performances when compared to it's Matlab version. This is probably due to quantification issues and the low-precision floating-point arithmetic computation within the FPGA. 






%

\subsection{Real data scenario}
\label{ssection:exp-real}
The real data scenario test has been carried out within the campus of ISAE-SUPAERO. The  equipment was mounted on the rooftop of a car. To this end, we used a USRP-X310 (Universal Software Radio Peripheral) from Ettus in order to acquire the combined 8-bit in-phase and quadrature samples of the GNSS signal at 4 MHz. A positioning reference system was set up consisting of two NovAtel PwrPak7-E1 GNSS+IMU and a dedicated Waypoint offline post-processing software, enabling DGNSS (Differential GNSS) corrections. The IMU data used within this test are provided by the mobile station at 125 Hz. 


\begin{figure}[t]
    \centering
    \includegraphics[width=0.4\textwidth]{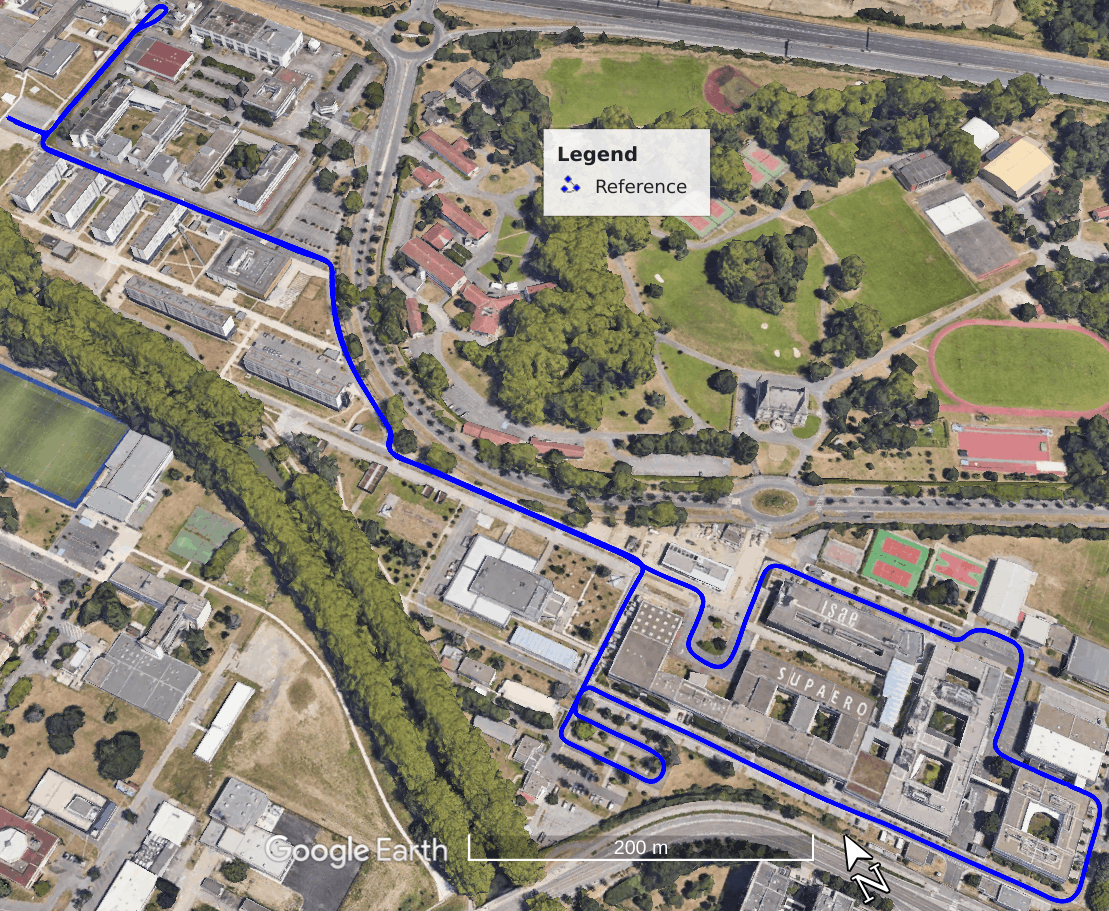}
    \caption{\textbf{ISAE-SUPAERO campus trajectory visualized with Google Earth.}}
    \label{fig:campus-traj}
\end{figure}

\begin{figure}
     \centering
     \begin{subfigure}[b]{0.48\textwidth}
         \centering
         \includegraphics[width = \textwidth]{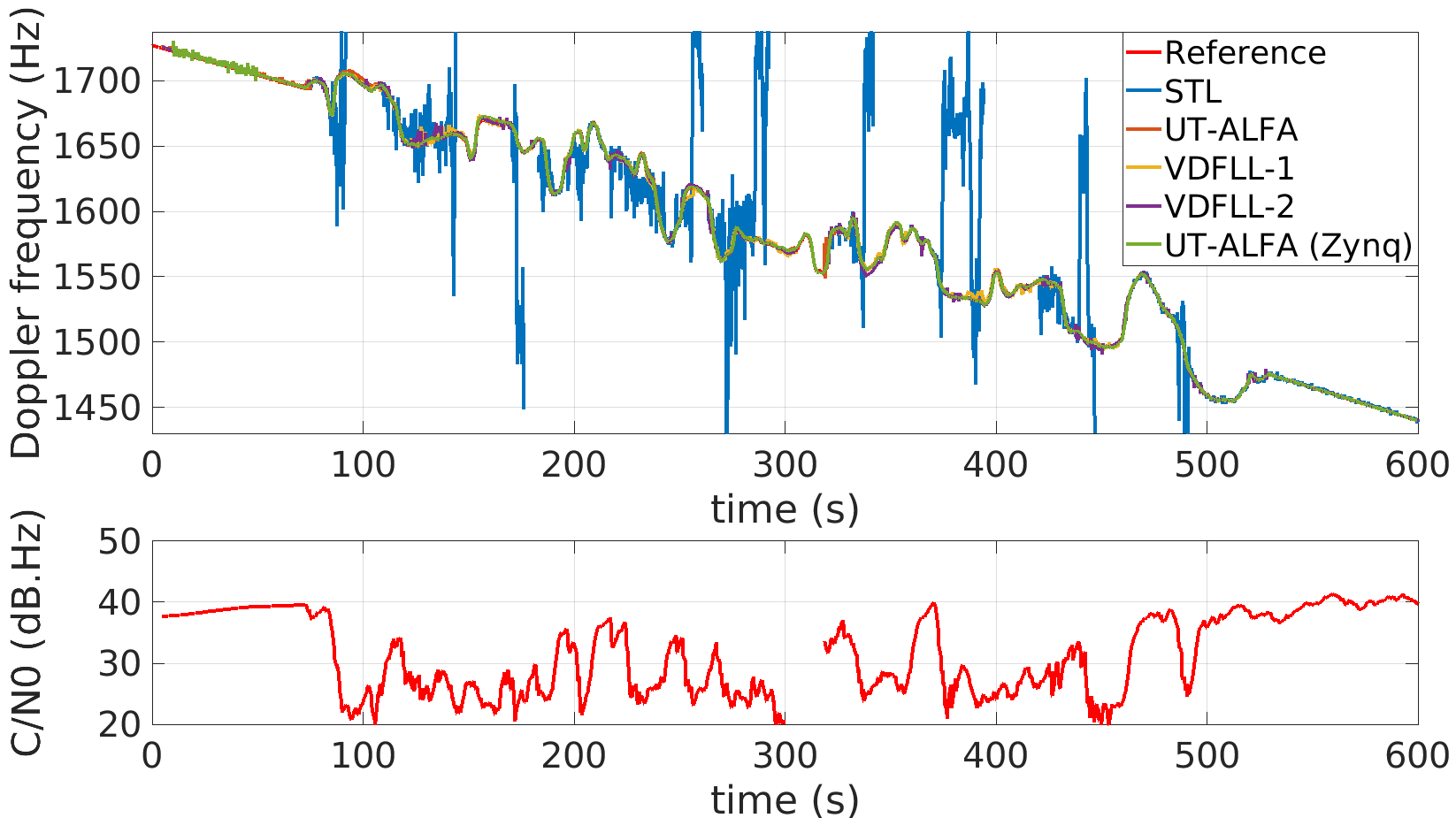}
         \caption{}
         \label{fig:realdata-gps5}
     \end{subfigure}
     \hfill
     \begin{subfigure}[b]{0.48\textwidth}
         \centering
         \includegraphics[width = \textwidth]{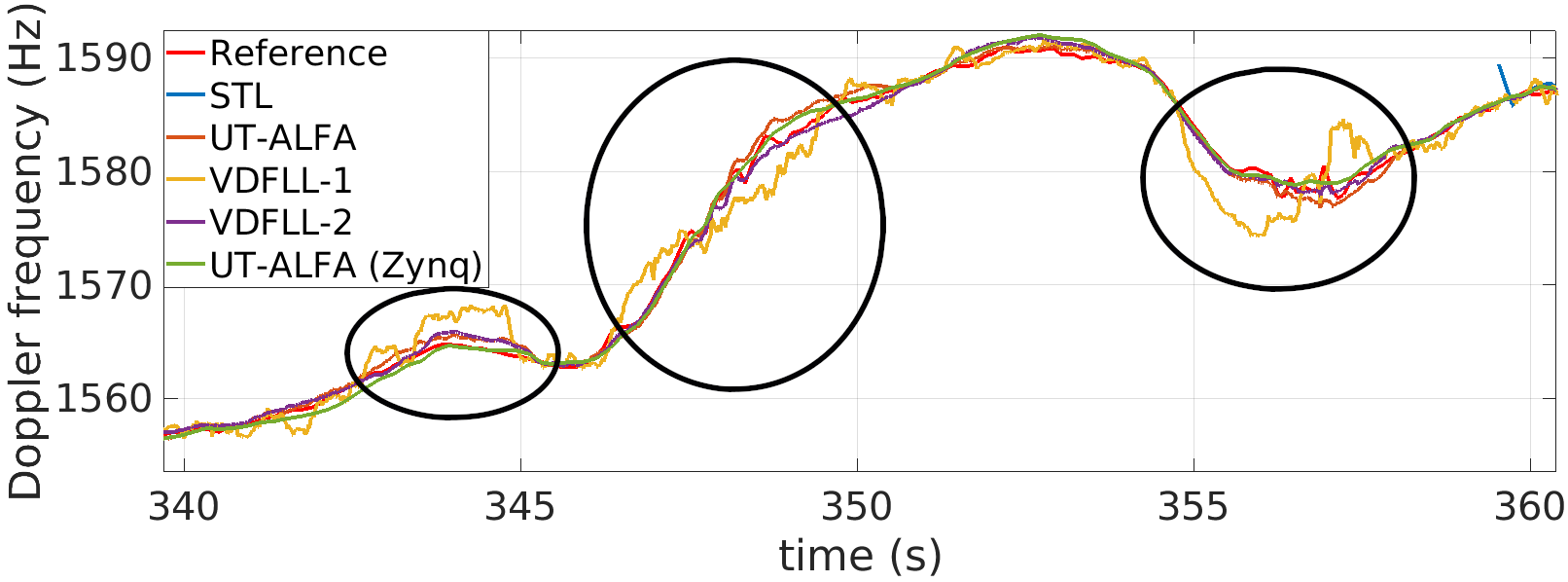}
        \caption{}
        \label{fig:realdata-gps5-zoom}
     \end{subfigure}
     \caption{\textbf{(a) Doppler frequency and estimated C/N0 for GPS satellite 5, (b) zoom on Doppler frequency at time range 340-360 seconds.}}
     \label{fig:realdata-gps}
\end{figure}

\begin{figure}
     \centering
     \begin{subfigure}[b]{0.48\textwidth}
         \centering
         \includegraphics[width = \textwidth]{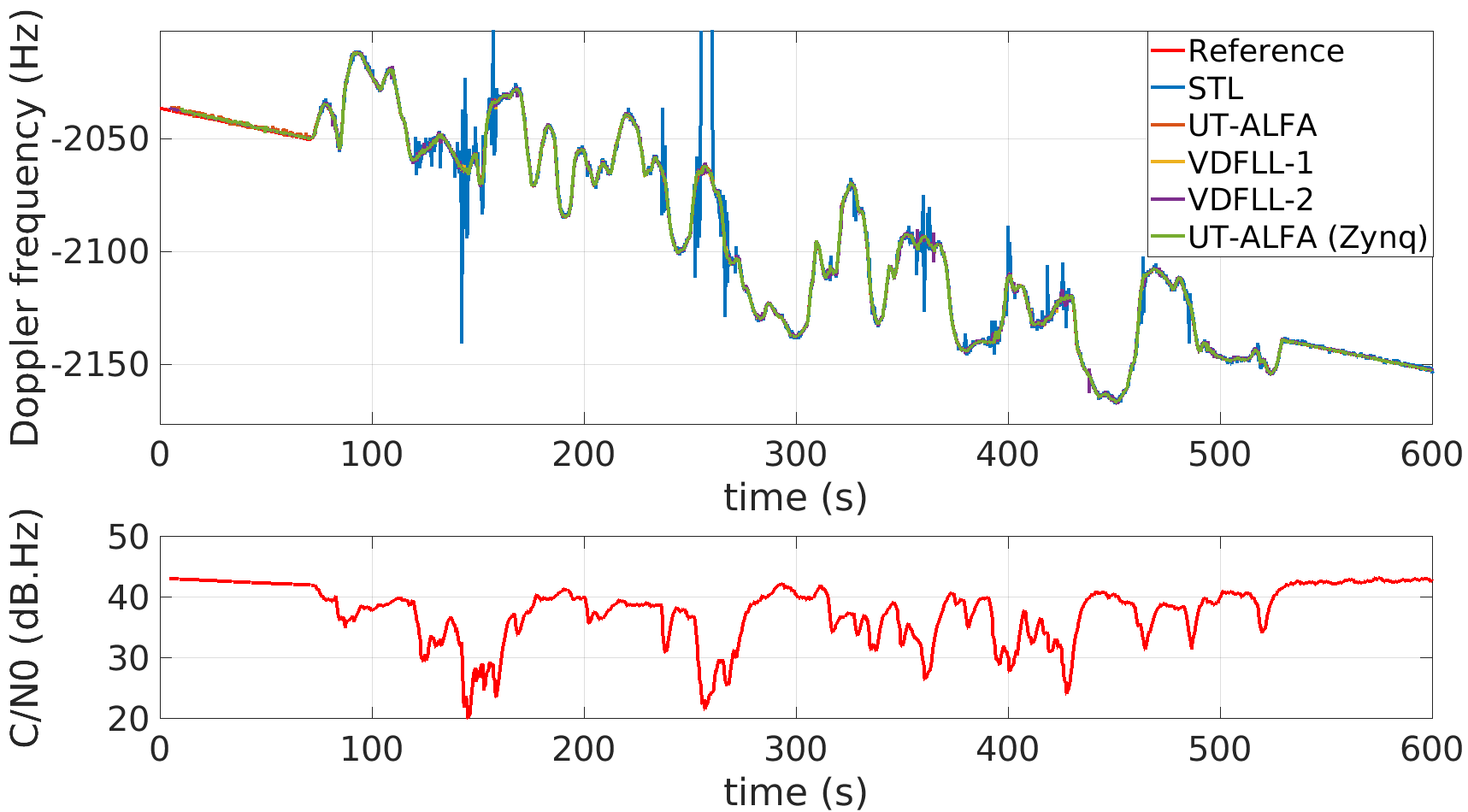}
         \caption{}
         \label{fig:realdata-gal21}
     \end{subfigure}
     \hfill
     \begin{subfigure}[b]{0.48\textwidth}
         \centering
         \includegraphics[width = \textwidth]{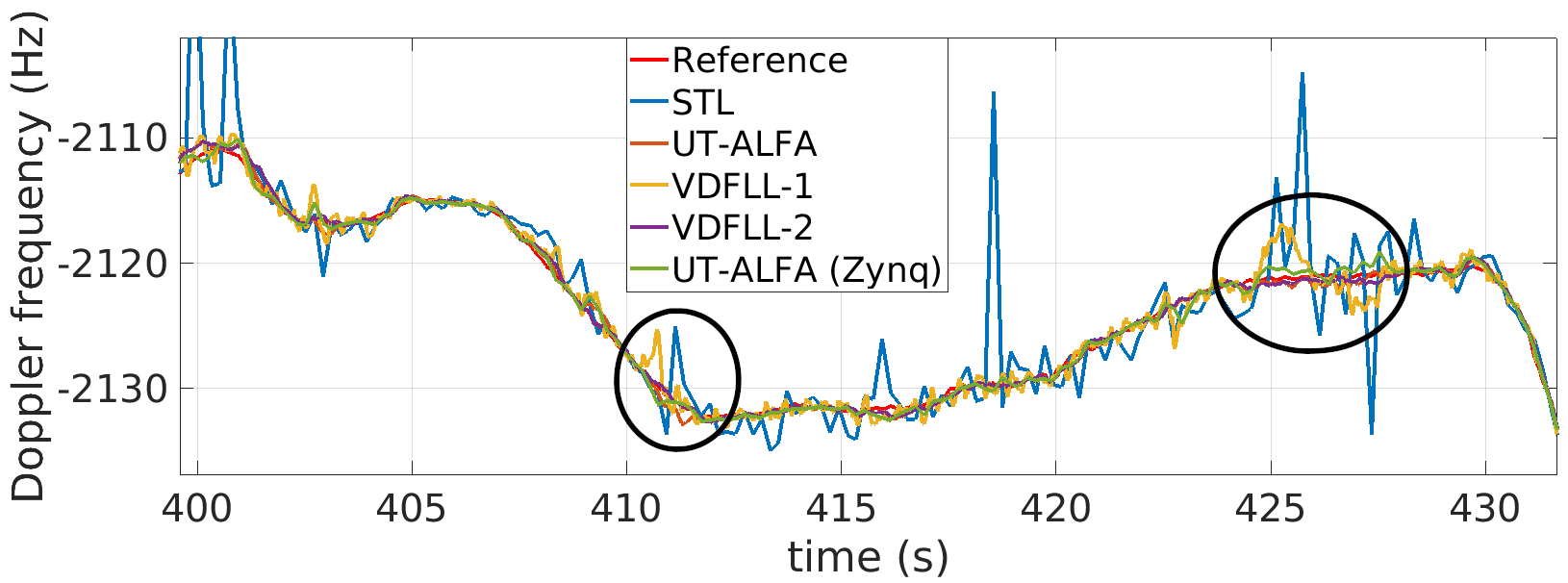}
        \caption{}
        \label{fig:realdata-gal21-zoom}
     \end{subfigure}
     \caption{\textbf{(a) Doppler frequency and estimated C/N0 for Galileo satellite 21, (b) zoom on Doppler frequency at time range 400-430 seconds.}}
     \label{fig:realdata-gal}
\end{figure}




Results are shown for the different architectures in Fig. \ref{fig:realdata-gps} and \ref{fig:realdata-gal}, respectively for a GPS and a Galileo satellite. Fig. \ref{fig:realdata-gps5-zoom} and \ref{fig:realdata-gal21-zoom} are zooms of Fig. \ref{fig:realdata-gps5} and \ref{fig:realdata-gal21} for a specific time range. Notice that when the C/N0 drops to 20 dB.Hz, the STL channels loose track, which is not the case for the VTL channels. Moreover, as expected, the STL channels are strongly impacted by C/N0 values less than 35 dB.Hz, whereas the VTL channels are not. Furthermore, when comparing the VDFLLs with the UT-ALFA (Matlab and on Zynq), the latter shows similar, or even better, performances in some cases. Indeed, as shown in Fig. \ref{fig:realdata-gps5-zoom} and \ref{fig:realdata-gal21-zoom}, the VDFLL-1 tends to degrade slightly for C/N0 values less than 30 dB.Hz, as highlighted by the black ellipses. Besides, the VDFLL-1 has additional noise in the tracking channels while the UT-ALFA presents smoother Doppler frequency estimates and is less effected by weak C/N0 values. Finally, the UT-ALFA on Zynq has equivalent performances compared to the UT-ALFA implemented in Matlab, showing the applicability of the proposed architecture.

\section{Conclusion}

This paper presents an alternative to standard GNSS receivers VTL architectures, which are ultra-tightly coupled with an INS. The main objective of our study was to propose a trade-off between performance and ease of implementation, while targeting FPGA components. This resulted in the UT-ALFA where the PLL bandwidth of the tracking loops are adapted according to the Doppler frequency rate calculated by the navigation filter. The UT-ALFA was realized on Matlab and on a Xilinx UltraScale. The latter showed good real-time performances and highlighted the simplicity of the implementation compared to a VDFLL architecture. Future work includes to implement the navigation filter on the ARM processor of the Zynq as well as extending the study on improving code delay. 


\bibliographystyle{ieeetr} 
\bibliography{gnss-bibliography}

\section{Glossary}
    \begin{tabular}{ll}
        \textit{C/N0}: & Carrier-to-noise density ratio \\
        \textit{EKF}: & Extended Kalman Filter \\
        \textit{FE}: & Front-End \\
        \textit{FPGA}: & Field-Programmable Gate Arrays\\
        \textit{GNSS}: & Global Navigation Satellite System \\
        \textit{GPS}: & Global Positioning System \\
        \textit{HDL}: & Hardware Description Language \\
        \textit{IMU}: &  Inertial Measurement Unit\\
        \textit{INS}: &  Inertial Navigation System\\
        \textit{IP}: & Intellectual Property \\
        \textit{KF}: & Kalman Filter \\
        \textit{NCO}: & Numerically Controlled Oscillator \\
        \textit{PLL}: & Phase Lock Loop \\
        \textit{PRN}: & Pseudo Random Noise \\
        \textit{SoC}: & System on Chip\\
        \textit{STL}: &  Scalar Tracking Loop\\
        \textit{VDFLL}: & Vector Delay and Frequency Lock Loop \\
        \textit{VDLL}: & Vector Delay Lock Loop \\
        \textit{VTL}: & Vector Tracking Loop
    \end{tabular}

\end{document}